\documentstyle[12pt,aps,prl]{revtex}
\tighten
\begin{document}
\title {Null weak singularities in plane-symmetric spacetimes} 
\author{Amos Ori}
\address{
Department of Physics, Technion---Israel Institute of Technology,
32000 Haifa, Israel.}
\date{\today}

\maketitle

\begin{abstract}

We construct a new class of plane-symmetric solutions possessing a 
curvature singularity which is null and weak, like the spacetime 
singularity at the Cauchy horizon of spinning (or charged) black holes. 
We then analyse the stability of this singularity using a rigorous non-perturbative 
method. We find that within the framework of (linearly-polarized) 
plane-symmetric spacetimes this type of null weak 
singularity is locally stable. Generically, the singularity is also 
scalar-curvature. These observations support the new picture of the 
null weak singularity inside spinning (or charged) black holes, which is 
so far established primarily on the perturbative approach.
\end{abstract}

\section{Introduction} \label{sec 1}

The Kerr solution \cite{1} represents the geometry of the general 
stationary, vacuum, spinning black hole (BH). This solution has an 
obvious relevance to reality, because realistic astrophysical BHs are 
generally spinning \cite{2}. It is well known that a Kerr BH has a 
Cauchy horizon (CH) - a null hypersurface which marks the boundary of 
the domain of dependence for any initial hypersurface in the external 
world. A similar CH also exists in the Reissner-Nordstrom (RN) 
geometry, which describes spherically-symmetric charged BHs. The 
presence of a CH inside a spinning or charged BH is disturbing, because 
the Einstein equations do not provide a unique prediction for the 
extension of the geometry beyond the CH. However, it is also well 
known that the CH of RN and Kerr is unstable; Namely, linear 
perturbations of various types develop singularities at the CH, 
\cite{3}--\cite{8} suggesting that in a generic situation the smooth CH of 
Kerr or RN will be replaced by a curvature singularity.
	
For many years, the nature and exact location of this singularity were 
not completely clear. The prevailing point of view in the last few 
decades was that the spacetime singularity inside BHs is of the BKL 
\cite{9} type, i.e. spacelike, oscillatory, and tidally-destructive. In the 
last few years, however, there is steadily growing evidence that a 
singularity of a completely different character forms at the CH of 
spinning \cite{10,11} and charged \cite{12}--\cite{15} BHs. This 
singularity is 
null \cite{10,12,13,16} rather than spacelike, and weak \cite{10,13} 
(in Tepler's terminology \cite{17}), rather than tidally-destructive.
	
So far, the most direct evidence for the formation of a null weak 
singularity inside generic spinning BHs stems from the nonlinear 
perturbation analysis of the interior Kerr geometry \cite{10,11}. It 
would obviously be important to confirm the perturbative results from 
alternative, non-perturbative, directions of research. Motivated by this, 
Yurtsever \cite{18} suggested that plane-symmetric spacetimes could 
serve as an excellent test-bed for further exploring and testing the 
new picture of the BHs' null weak singularity, emerging from the 
perturbative approach. Yurtsever based his suggestion on the following 
argument: If indeed a null curvature singularity exists at the CH of 
generic spinning (or charged) black holes, there should exist 
corresponding plane-wave solutions \cite{19,20} which admit a 
(locally) similar type of null singularity. Therefore, understanding the 
rule of such null singularities in plane-wave and colliding plane-waves 
(CPW) solutions \cite{21,22} may provide important insight into the 
issue of stability of the null CH singularity inside spinning (or charged) 
BHs.
	
In this paper we shall attempt to undertake this goal. We shall first 
construct an exact linearly-polarized (LP) ingoing plane-wave solution 
which admits a weak curvature singularity on a null hypersurface. Then 
we shall analyse the stability of this type of singularity, within the 
framework of (LP) plane-symmetric spacetimes, by a rigorous non-perturbative 
method: First we shall demonstrate the stability with 
respect to generic {\it ingoing} plane-wave perturbations (which preserve 
the plane-wave character of the solution). Then we shall introduce 
{\it outgoing} plane-symmetric perturbations as well, and analyse their 
effect on the structure of the singularity. The outgoing perturbations 
convert the geometry into a (LP) CPW spacetime. We shall show that 
the singularity remains null and weak, though curvature scalars, which 
were strictly zero in the original plane-wave solution, generically 
blow up when the outgoing perturbations turn on. (This situation is 
fully analogous to what is known about the black holes' CH singularity: 
In a spherical charged BH, if the radiation is purely ingoing the CH 
singularity is non-scalar \cite{16}, but when one adds outgoing 
radiation, the singularity becomes scalar-curvature - the mass-inflation 
singularity \cite{12}. In the case of a spinning vacuum BH, the 
CH singularity is scalar-curvature. \cite{10,23}) We conclude that the 
null weak singularity is locally stable within both frameworks of 
plane-wave solutions and CPW solutions (though the non-scalar 
character of the null singularity in plane-wave solutions is unstable 
within the framework of CPW solutions). This provides a strong support 
to the above-mentioned new picture of the BHs' singularity.
	
In Ref. \cite{18} Yurtsever presented a certain limiting process which 
maps a (local neighborhood of a) generic null singularity into a plane-wave 
null singularity. Based on this limiting process, Yurtsever 
correctly pointed out that if indeed a generic null singularity exists 
inside black holes, it should be asymptotically similar to a generic null 
singularity of a plane-wave solution. Yurtsever further argued that this 
plane-wave null singularity should coincide with the ``singular Cauchy 
horizon" of the plane-wave solution. But the Killing Cauchy horizons of 
plane-wave solutions are known to be non-generic and unstable within 
the context of global initial-value problem for CPW spacetimes. 
\cite{22} This led Yurtsever to the conclusion that a null singularity 
(e.g. inside black holes) must be locally unstable and hence unrealistic. 
This conclusion, we argue, is incorrect; Its derivation requires one to 
make two closely related assumptions:
\newline
{\it Assumption 1}:   The scenario of a formation of a black hole, with a null 
curvature singularity inside it, from regular initial data, can be 
approximated in a global sense by some plane-wave (or CPW) solution 
with a null singularity, which evolves from regular initial data.
\newline
{\it Assumption 2}:   Correspondingly, the latter null singularity should be 
located at the Killing Cauchy horizon of the plane-wave (or CPW) 
solution.
\newline
Both assumptions result from the confusion of local and global aspects 
of the problem. The statement that the black-hole's null singularity is 
well approximated by a plane-wave null singularity holds only on a 
{\it local} basis. This is obvious from the nature of the limiting process 
described in Ref. \cite{18}, which assumes a null geodesic $\gamma $ 
that intersects the null singularity: It is only the ``tubular" immediate 
neighborhood of $\gamma $ (near its intersection point with the 
singularity) which admits the approximate similarity to a plane-wave 
solutions. (See Fig. 1, in which this neighborhood is denoted by $N$.) 
Since the initial hypersurface of the black-hole spacetime is remote 
from the null singularity inside the BH, there is no way to extend $N$ 
up to this initial hypersurface. Therefore, despite the local similarity 
of the neighborhoods of the null singularities in the two spacetimes, 
their global features are totally different. In particular, there is no 
reason to relate the BH's null singularity to the plane-wave Killing 
Cauchy horizon. Moreover, when considered as a simplified approximate 
model for the BH's null singularity, the issue of stability of the plane-wave 
null singularity must be considered from the {\it local} point of view 
(see below), and not from the global one based on Cauchy evolution from 
regular initial data.

\begin{figure}
\input{epsf}
\centerline{\epsfysize 10.0cm
\epsfbox{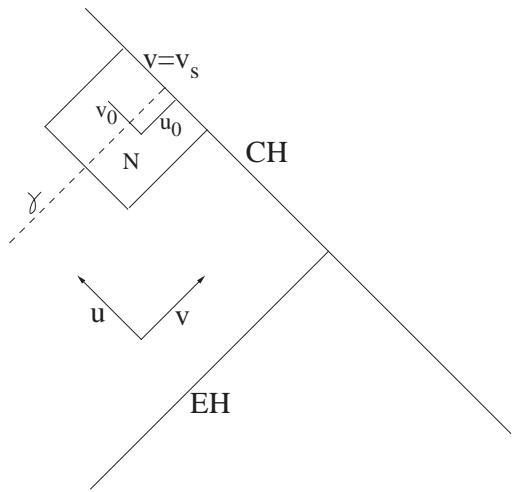}}
\caption{
The local initial-value setup for the plane-symmetric spacetime, 
and its relation to the inner structure of the black hole. EH and CH are 
the event horizon and the singular Cauchy horizon of the black hole, 
respectively. The two characteristic initial null hypersurfaces, 
$u=u_0$ and $v=v_0$, are located in the neighborhood $N$ of the 
intersection point of the CH singularity and the outgoing null geodesic 
$\gamma $.
}
\end{figure}
	
Our approach to the issue of local stability is based on a {\it local 
(characteristic) initial-value setup}. Namely, we introduce a pair of 
characteristic initial null hypersurfaces (denoted $u_0$ and $v_0$ in 
Fig. 1), both located inside the small local neighborhood $N$ of a point 
on the null singularity. In such an initial-value setup, the generic CPW 
solution depends on two freely-specifiable functions, one of the 
ingoing null coordinate $v$ and one of the outgoing null coordinate $u$. 
(In the special case of a plane-wave solution, one of these functions 
degenerates to a constant.) We shall first construct a specific ingoing 
plane-wave solution which admits a null curvature singularity at  
$v=v_s$. Then we shall add perturbations to both the above mentioned 
initial functions. These perturbations are generic, though bounded (and 
restricted in some weak sense). We shall show, by means of a non-
perturbative analysis, that the null singularity at $v=v_s$
 survives these perturbations. That is, within the framework of CPW 
spacetimes, the null singularity is locally stable.
	
In this local setup, the outgoing initial hypersurface $u_0$ necessarily 
intersects the null singularity (obviously, if we chop this hypersurface 
before $v=v_s$, the null singularity will not be included in the domain 
of dependence). In other words, in order to recover the null singularity, 
we must start with initial data at $u=u_0$ which are themselves 
singular at a certain point ($u=u_0$,$v=v_s$). \cite{24} (This singular 
point will then be the ``seed" of the null singularity, which will extend 
along the hypersurface $v=v_s$.) This disadvantage is unavoidable, 
because, as was explained above, the {\it global} plane-wave initial-value 
problem (starting from regular initial data) has little to do with the 
issue of null singularities inside black holes. In physical terms, what 
we show here is that a singularity (of a certain type) at the initial data 
generically propagates along the characteristic line and thereby 
produces a null singularity. This observation is not trivial: One might 
conceive that the nonlinearity of the field equations will generically 
prevent the formation of a null singularity, and a spacelike singularity 
will form instead (this is essentially the possibility suggested by 
Yurtsever \cite{18}). Our local stability analysis shows that this is not 
the case: The null singularity of plane waves is locally stable within 
the frameworks of plane-wave and CPW spacetimes.
	
Recently, Ori and Flanagan (OF) \cite{25} used another construction to 
demonstrate the local genericity of null weak singularities. The 
construction by OF is more powerful than the one presented here, as it 
demonstrates local (functional) genericity within the framework of the 
fully-generic class of vacuum solutions, without any symmetry. The 
present work is advantageous in certain respects, however: First, the 
mathematical construction by OF heavily depends on analyticity: It is 
restricted to analytic initial-value functions, and is primarily based on 
the Cauchy-Kowalewski theorem. The present construction does not 
make any assumptions about the analyticity of the initial functions (we 
only demand smoothness), and is based on the hyperbolic nature of the 
field equations. Second, the type of {\it local asymptotic behavior} covered 
by the present analysis is slightly more general than that of OF, as 
explained in section \ref{sec5}. In addition, the generalization to null 
singularities with more realistic types of local asymptotic behavior 
will be much easier to implement within the present framework of CPW 
spacetimes (this is again a consequence of the lack of any reference to 
analyticity in the present method; see section \ref{sec5}). Third, the simplicity 
of the CPW solutions makes the present construction more useful for 
investigating various features of the null singularity. We should also 
mention another attempt to use a local construction for investigating 
the BH's null CH singularity, made earlier by Brady and Chambers 
\cite{26}. Their construction, however, only addressed the constraint 
equations on two null hypersurfaces, and the compatibility with the six 
evolution equations was not considered there. In particular, the 
analysis in Ref. \cite{26} does not rule out the possibility that a 
spacelike singularity will form immediately at the singular ``point" 
[analogous to ($u_0$,$v_s$)] at the future edge of the outgoing initial 
null hypersurface. The present construction (like that of OF) takes care 
of the full set of vacuum Einstein equations.
	
Throughout this paper we restrict attention to {\it linearly-polarized} (LP) 
plane-wave and CPW solutions, for simplicity. We do not expect the 
qualitative results to be different in the more general case of 
arbitrarily polarized solutions. The analysis in the arbitrarily polarized 
case is more complicated, but still appears to be feasible.
	
This paper is organized as follows. In section \ref{sec 2} we shall construct an 
explicit ingoing LP plane-wave solution with a non-scalar parallelly-propagated (PP) 
null weak singularity, to which we shall refer as the 
{\it basic solution}. Then, in section \ref{sec 3} we shall perturb this basic solution 
by generic (though bounded) LP {\it ingoing} perturbations, and show that the 
singularity remains non-scalar, PP, null, and weak. This demonstrates 
that the non-scalar null weak singularity is a generic feature of (LP) 
plane-wave spacetimes. In section \ref{sec 4} we shall add generic (LP) 
perturbations in the {\it outgoing} direction. The geometry is now described 
by the CPW solution. We shall show that the singularity remains null 
and weak, though generically the outgoing perturbations convert the 
non-scalar PP curvature singularity into a scalar-curvature singularity. 
Finally, in section \ref{sec5} we shall discuss the extent and implications of 
our results.

\section{Basic  Singular  Plane-Wave  Solution} \label{sec 2}

The LP plane-wave spacetimes can be described by the line element
$$ds^2=-du\,dv+e^{-U(v)}\,\left[ {e^{V(v)}dx^2+e^{-V(v)}dy^2} 
\right]_{}\;.\eqno(1)$$ 
The only non-trivial vacuum Einstein equation is
$$2U_{,vv}-{U_{,v}}^2-{V_{,v}}^2=0\;.\eqno(2)$$
Since there are two non-trivial functions [$U(v)$ and $V(v)$] and one 
constraint, Eq. (2), plane-wave solutions are described by one arbitrary 
function of $v$. To simplify the calculations, we take the freely-specified 
function to be $U(v)$. Equation (2) then becomes an integral 
for $V(v)$. [When the freely-specified function is taken to be $V(v)$, 
Eq. (2) becomes a non-linear differential equation for $U(v)$.]
	
We wish to construct an explicit solution in the range $v_0\le v<v_s$ 
which develops a null weak curvature singularity at $v=v_s$. In 
general, such a solution will be obtained from any function $V(v)$ 
which is smooth at $v_0\le v<v_s$ and continuous at $v=v_s$, but with 
$V_{,v}$ diverging at $v=v_s$. We shall now construct a simple explicit 
solution of this type. To shorten the notation, we take $v_s=0$ and 
define
$$w\equiv |v|^{1/ n}=(-v)^{1/ n}\;,\eqno(3)$$
where $n$ is any constant $>2$ . (Note that with the choice $v_s=0$, 
both $v$ and $v_0$ are negative.) We now take
$$U(v)=-\,(n-2)\,a^2w^2\equiv U_0(v)\;,\eqno(4)$$ 
where $a$ is a positive dimensional constant. Solving Eq. (2), we obtain
$$V(v)=(n-2)\,\left[ {aw\,\left( {1-a^2\,w^2} \right)^{1/ 2}+\arcsin 
\left( {aw} \right)} \right]\equiv V_0(v)\;.\eqno(5)$$
(For concreteness we took here the positive root for $V_{,v}$, and set 
the integration constant such that $V$ vanishes at $v=0$. This causes 
no loss of generality: Adding a constant to $V$ or changing its sign 
amounts to rescaling the coordinates $x$ and $y$ or interchanging 
them, respectively.) In order to ensure the validity and regularity of 
$V(v)$ in the entire range $v_0\le v<0$, we take $0<a\,<|v_0|^{-1/ n}$, 
so that $0<aw<1$. 
	
We shall refer to the explicit plane-wave solution (4,5) as the {\it basic 
solution}. In terms of our characteristic initial-value setup, this plane 
wave solution may be viewed as evolving from the initial data (4,5) at 
the characteristic hypersurface $u=u_0$, with trivial (i.e. u-
independent) initial data along the other hypersurface, $v=v_0$.
	
An expansion of Eq. (5) near $v=0$ yields
$$V=2(n-2)aw+O\,(w^3)\;.\eqno(6)$$
For any function $F(w)$, we find
$$F_{,v}=-n^{-1}|v|^{1/ n-1}f_{,w}\quad ,\quad F_{,vv}=n^{-2}\left[ 
{|v|^{2/ n-2}f_{,ww}-(n-1)|v|^{1/ n-2}f_{,w}} \right]\;.\eqno(7)$$
Therefore, at $v=0$ the derivatives of $V$ and $U$ with respect to $v$ 
diverge: 
$$V_{,v}\,\cong -\,\left[ {2n^{-1}(n-2)a} \right]\;\,|v|^{1/ n-
1}\;\;,\;\;U_{,v}\,\propto \,v^{2/ n-1}\;.\eqno(8)$$
(This divergence has an invariant meaning, because $v$ is the affine 
parameter for the null geodesics of constant $x,y,u$ .) The second-order 
derivatives diverge even faster:
$$V_{,vv}\,\cong -\,\left[ {2n^{-2}(n-1)(n-2)a} \right]\;\,|v|^{1/ n-
2}\;\;,\;\;U_{,vv}\,\propto \,v^{2/ n-2}\;.\eqno(9)$$
As a consequence, various components of the Riemann tensor diverge at 
$v=0$. For example, one finds that 
\footnote {This expression is 
obtained from a direct calculation of the Weyl tensor, which is equal to 
the Riemann tensor in the vacuum spacetimes considered here.} 
\footnote {The most divergent components of Riemann are $R_{vxvx}$ 
and $R_{vyvy}$ ($R_{vxvy}$ vanishes identically). The expression for 
$R_{vyvy}$ is the same as that of $R_{vxvx}$, except that $V$ is 
replaced by $-V$. (The same relation holds for the corresponding PP 
components discussed below.)}
$$R_{vxvx}={1 \over 2}e^{V-U}\left( {-V_{,vv}+U_{,v}V_{,v}} 
\right)\;,\eqno(10)$$
and thus near $v=0$,
$$R_{vxvx}\cong -{1 \over 2}V_{,vv}\cong n^{-2}(n-1)(n-2)a|v|^{1/ n-
2}\;.$$  Since all curvature scalars vanish in plane-wave spacetimes, 
there is no scalar-curvature singularity at $v=0$. One can easily verify, 
however, that parallelly-propagated (PP) components of the Riemann 
tensor diverge along null and timelike geodesics intersecting $v=0$. For 
example, along a null geodesic of constant $x,y,u,$ a convenient PP 
tetrad is
$$e_x^\alpha =e^{(U-V)/ 2}\kern 1pt \delta _x^\alpha \;\;,\;\;e_y^\alpha 
=e^{(U+V)/ 2}\kern 1pt \delta _y^\alpha \;\;,\;\;e_u^\alpha =\sqrt 
2\,\delta _u^\alpha \;\;,\;\;e_v^\alpha =\sqrt 2\,\kern 1pt \delta 
_v^\alpha \;.\eqno(11)$$
The PP tetrad component
$$e_v^\alpha \,e_x^\beta \,e_v^\gamma \,e_x^\delta \,R_{\alpha \beta 
\gamma \delta }=2e^{(U-V)}\kern 1pt R_{vxvx}$$
diverges like $|v|^{1/ n-2}$.
	
Thus, there is a PP curvature singularity at $v=0$. However, because 
$U=V=0$ there, the line element (1) is well-define (i.e. finite and non-degenerate) 
even at $v=0$. It then follows that the curvature 
singularity at $v=0$ is weak (in Tipler¹s terminology \cite{17}). We 
conclude that the hypersurface $v=0$ is a null, weak, PP curvature 
singularity. Note also that no other singularity occurs in the range 
$v_0\le v<0$ , because $U$ and $V$ are smooth ($C^\infty $) functions 
of $w$.

\section{Stability To Ingoing  Plane-Wave  Perturbations} \label{sec 3}

In this section we shall analyse the stability of the basic plane-wave 
solution (4,5) within the framework of (LP) plane waves. The basic 
solution is characterized by the initial functions $U_0(v)$ and 
$V_0(v)$, Eqs. (4,5). We shall now add a small (though finite) 
perturbation $\Delta U(v)$ to the initial function $U_0(v)$:
$$U(v)=U_0(v)+\Delta U(v)\;.\eqno(12)$$
In terms of the characteristic initial-value setup, this amounts to 
perturbing the initial data at the characteristic hypersurface $u=u_0$, 
while leaving the trivial ($u$-independent) initial data at $v=v_0$ 
unchanged (apart from a trivial constant shift, to allow for continuity 
at the intersection point of the two initial hypersurfaces). We assume 
that the perturbation $\Delta U(v)$ is a smooth function of $v$ in the 
range $v_0\le v\le 0$. By virtue of the constraint equation (2), this 
perturbation of $U(v)$ will lead to a corresponding perturbation in 
$V(v)$, which we now analyse. We first write Eq. (2) as
$${V_{,v}}^2=2U_{,vv}-{U_{,v}}^2={V_{0,v}}^2+\left[ {2\Delta U_{,vv}-
{\Delta U_{,v}}^2-2U_{0,v}\Delta U_{,v}} \right]\;.$$
Converting the derivatives of $V$, $V_0$, and $U_0$ (but not of $\Delta 
U$) from $v$ to $w$, recalling $v_{,w}=-nw^{n-1}$, and 
selecting the positive root (as before), we obtain an expression of the 
form
$$V_{,w}=2a(n-2)\left[ {1-a^2w^2+b_1\Delta U_{,v}\,w^n+({b_2\Delta 
U_{,v}}^2+b_3\Delta U_{,vv})w^{2n-2}} \right]^{1/ 2}\;,\eqno(13)$$
where $b_1$, $b_2$ and $b_3$ are constants that depend on $a$ and 
$n$. In the basic solution ($\Delta U=0$), the term in brackets is 
strictly positive in $v_0\le v\le 0$ (recall that $|aw|<1$). Therefore, 
bounds $B_1$,$B_2$ (which may depend on $a$, $n$, and $v_0$) exist 
such that the term in brackets will be strictly positive in $v_0\le v\le 
0$ for any $\Delta U(v)$ satisfying 
$$\left| {\Delta U_{,v}} \right|<B_1\;\,,\;\,\left| {\Delta U_{,vv}} 
\right|<B_2 \eqno(14)$$
throughout this range. For all perturbations satisfying the inequalities 
(14), $V_{,w}$ is well-defined and smooth in $v_0\le v\le 0$, and so is 
$V(w)$.
	
Turning back from $w$ to $v$, we first observe that both $U(v)$ and 
$V(v)$ are smooth in $v_0\le v<0$ (and finite at $v=0$). That is, no new 
singularity appears in the range $v<0$. We still need to check the effect 
of the perturbation on the features of the singularity at $v=0$. From Eq. 
(13) it is obvious that both $V_{,v}$ and $V_{,vv}$ are unaffected at the 
leading order in $w$, so the asymptotic behavior at $v=0$ is still 
correctly described by Eqs. (8,9). It then follows that PP Riemann 
components [with respect to the tetrad (11)] diverge just as in the 
basic solution, e.g.
$$e_v^\alpha \,e_x^\beta \,e_v^\gamma \,e_x^\delta \,R_{\alpha \beta 
\gamma \delta }=2e^{(U-V)}\kern 1pt R_{vxvx}\propto \;|v|^{1/ n-2}\;.$$ 
(However, all curvature scalars vanish, as we are still dealing with a 
plane-wave solution.) The hypersurface $v=0$ thus remains a non-scalar PP 
curvature singularity. Since both $U$ and $V$ are finite at 
$v=0$, the singularity is weak. We conclude that the non-scalar null 
weak PP curvature singularity of the basic solution is stable to small 
(but generic) ingoing plane-wave perturbations of the type considered 
here.

\section{Stability  To  Outgoing  Plane-Symmetric  Perturbations} 
\label{sec 4}

Next, we check stability with respect to outgoing perturbations; 
namely, we shall now assume that non-trivial initial data are present 
also on the ingoing initial hypersurface $v=v_0$. The geometry is still 
plane-symmetric, but is no longer a plane-wave; Rather, it is described 
by the (LP) CPW solution. The line element is
$$ds^2=-e^{-M}du\,dv+e^{-U}\,\left[ {e^Vdx^2+e^{-V}dy^2} 
\right]_{}\;,\eqno(15)$$
where the functions $M,U,V$ generally depend on both $v$ and $u$. This 
line element and the corresponding field equations have been discussed 
by several authors \cite{21,22}; Here we shall briefly present the field 
equations in a form suitable for our analysis (basically the same form 
as in Ref. \cite{22}). The vacuum Einstein equations include five non-trivial 
equations,
$$U_{,uv}-U_{,u}U_{,v}=0\eqno(16a)$$
$$V_{,uv}-{\textstyle{1 \over 
2}}(U_{,u}V_{,v}+V_{,u}U_{,v})=0\eqno(16b)$$
$$M_{,uv}-{\textstyle{1 \over 2}}(V_{,u}V_{,v}-U_{,u}U_{,v})=0 
\eqno(16c)$$  	
and
$$2\,U_{,vv}-{U_{,v}}^2-{V_{,v}}^2+2M_{,v}U_{,v}=0 \eqno(17a)$$
$$2\,U_{,uu}-{U_{,u}}^2-{V_{,u}}^2+2M_{,u}U_{,u}=0 \;.\eqno(17b)$$ 
The evolution equations (16a-c) form a closed system of hyperbolic 
equations for the three unknowns $U$, $V$, and $M$. In the 
characteristic initial-value setup, the values of these three functions 
are to be specified on the two characteristic initial null hypersurfaces, 
$u=u_0$ and $v=v_0$ (see Fig. 2). We shall denote the initial values of 
$U$, $V$, and $M$ on $u=u_0$ by $^vU(v)$, $^vV(v)$, and $^vM(v)$, 
correspondingly, and those on the other null hypersurface, $v=v_0$ , by 
$^uU(u)$, $^uV(u)$, and $^uM(u)$. We shall fix the gauge by demanding
$$^uM(u)=0\;\;,\;\;^vM(v)=0\;.\eqno(18)$$
Any solution of the evolution equations (16a-c) will satisfy the 
equations (17a,b), provided that the latter equations are satisfied on 
the two initial null hypersurfaces. Thus, in this gauge the Einstein 
equations are reduced to a set of three evolution equations (16a-c), 
supplemented by the demand that the six initial functions $^vV(v)$, 
$^vU(v)$, $^vM(v)$ and $^uV(u)$, $^uU(u)$, $^uM(u)$ will satisfy Eq. (18) 
and the two ordinary differential equations,
$$2\,^vU_{,vv}-{^vU_{,v}}^2-{^vV_{,v}}^2=0 \eqno(19a)$$
$$2\,^uU_{,uu}-{^uU_{,u}}^2-{^uV_{,u}}^2=0\;.\eqno(19b)$$ 
Correspondingly, a (LP) CPW spacetime is characterized by two 
arbitrary initial functions, $^vU(v)$ and $^uU(u)$; The two other 
nontrivial initial functions, $^vV(v)$ and $^uV(u)$, will be determined 
from Eqs. (19a,b).

\begin{figure}
\input{epsf}
\centerline{\epsfysize 10.0cm
\epsfbox{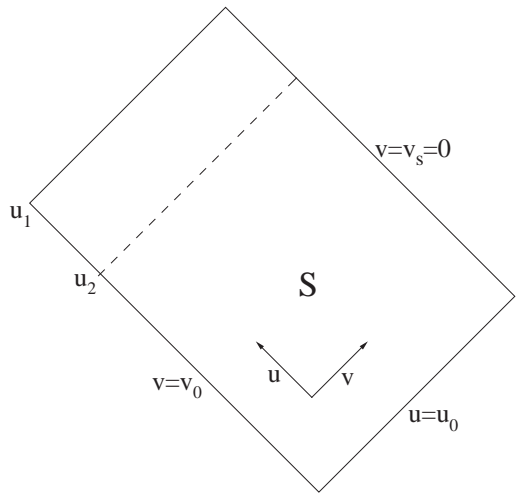}}
\caption{
The characteristic initial-value setup for the CPW solutions 
constructed in Sec. IV.
}
\end{figure}

In the class of CPW solutions, the ingoing plane-wave solutions form a 
subclass of measure zero, characterized by a trivial initial function 
$^uU(u)=const$. One then immediately obtains from Eq.(19b) and 
Eqs.(16a-c) that $U(u,v)=\kern 1pt {}^vU(v)$, $V(u,v)=\kern 1pt 
{}^vV(v)$, and $M(u,v)=0$, and Eqs. (1,2) are recovered. The plane-wave 
solutions considered in sections II and III correspond to 
$^vU(v)=U_0(v)$ and $^vU(v)=U_0(v)+\Delta U(v)$, respectively (and 
constant $^uU$). Here we shall consider nontrivial initial functions on 
both initial null hypersurfaces. On the outgoing hypersurface $u=u_0$, 
we take $^vU(v)=U_0(v)+\Delta U(v)$ as in section \ref{sec 3}. On the ingoing 
hypersurface $v=v_0$, the initial function $^uU(u)$ can be any smooth 
function of $u$ in the range $u_0\le u\le u_1$ (for some $u_1>u_0$) 
which satisfies the following two obvious requirements: 
\newline
(i)   $2\kern 1pt {}^uU_{,uu}-\kern 1pt {}{^uU_{,u}}^2>0$ in $u_0\le u\le 
u_1$; This will ensure the existence and smoothness of the function 
$^uV(u)$ [determined from Eq. (19b)] on the initial hypersurface 
$v=v_0$, throughout the relevant range $u_0\le u\le u_1$.
\newline
(ii)   $^uU(u=u_0)=\kern 1pt ^vU(v=v_0)$, which is dictated by 
continuity of $U(u,v)$ at the intersection point $(u_0,v_0)$.
\newline
Our goal is now to analyze the evolution of geometry inside the domain 
of dependence, $\{u_0\le u\le u_1\,\,,\,\,v_0\le v\le 0\}$.
	
We start by analyzing $U(u,v)$. Defining 
$$W=e^{-U}\;,\eqno(20)$$
Eq. (16a) is reduced to $W_{,\,uv}=0$. The solution, subject to the above 
initial data, is
$$W(u,v)=^vW(v)+^uW(u)-^uW(u_0)\;,\eqno(21)$$
where $^{u,v}W\equiv \exp (-^{u,v}U)$ and
$$^0W\equiv {}^uW(u=u_0)={}^vW(v=v_0)\;.$$ 
From the smoothness of $^vU(w)$ and $^uU(u)$ it then follows that 
$W(u,w)$ is $C^\infty $ in the entire domain $u_0\le u\le 
u_1\,\,,\,\,v_0\le v\le 0$, and hence $W(u,v)$ is $C^\infty $ in $u_0\le 
u\le u_1\,\,,\,\,v_0\le v\le 0$ (but not at $v=0$; see below). Since 
$U=-\ln (W)$, the only possible singularity of $U$ at $v<0$ is at the 
points where $W$ vanishes. Obviously, 
$$W(u=u_0,v)=\kern 1pt ^vW(v)=\exp \left[ {-^vU(v)} \right]$$ 
is strictly positive at $v_0\le v\le 0$.
\footnote {Note that $^vU$ 
is finite at $v=0$, despite the divergence of $^vU_{,v}$ there.}
By continuity of $W(u,w)$, there exists a constant $u_2$, 
$u_0<u_2<u_1$, such that $W>0$ in the entire domain $S\equiv 
\{u_0\le u\le u_2\,\,,\,\,v_0\le v\le 0\}$ (see Fig. 2). Consequently, 
$U(u,w)$ is smooth in $S$, and $U(u,v)$ is continuous in $S$. In 
addition, $U(u,v)$ is $C^\infty $ in the domain $S^-\equiv \{u_0\le u\le 
u_2\,\,,\,\,v_0\le v<0\}$.
	
Before we analyze the evolution of $V$ and $M$ , we denote a 
remarkable feature of the evolution equations (16a-c): The form of 
these equations remain unchanged if we replace the independent 
variables $u$ and $v$ by new ones, $\bar u(u)$ and $\bar v(v)$ (with the 
unknowns $U$, $V$, and $M$ unchanged); The only modification required 
is to replace $\partial u$ by $\partial \bar u$ and $\partial v$ by 
$\partial \bar v$. We use this freedom to transform from $v$ to $w$. 
Consider first Eq. (16b) (with $v$ replaced by $w$). Since $U(u,w)$ is 
known from Eqs. (20,21), this is a hyperbolic linear equation for 
$V(u,w)$. Both its coefficients (i.e. $U_{,u}$ and $U_{,w}$) and its 
initial data [i.e. ${}^uV(u)$ and 
${}^vV(w)$] are $C^\infty $ functions of $u$ 
and/or $w$. Therefore, this equation has a unique smooth solution 
$V(u,w)$ throughout $S$. Finally, consider Eq. (16c) for $M(u,w)$. The 
second term in that equation, which includes derivatives of $V$ and $U$ 
with respect to $u$ and $w$, is smooth in $S$, and the initial data at 
both initial hypersurfaces are $M=0$. Consequently, $M(u,w)$ is smooth 
throughout $S$.
	
We conclude that all three functions $U(u,w)$, $V(u,w)$, and $M(u,w)$ 
are $C^\infty $ throughout $S$ (even at $w=0$). Returning now from 
$w$ to the original independent variable $v$, we find that $U(u,v)$, 
$V(u,v)$ and $M(u,v)$ are continuous throughout $S$ and, moreover, are 
smooth in $S^-$. However, these functions will generically fail to be 
smooth at $v=0$. To analyse this lack of smoothness, we use Eq. (7) to 
evaluate the $v$-derivatives of $U$, $V$ and $M$. The maximal possible 
divergence rate of the first-order $v$-derivatives is $|v|^{1/ n-1}$ (in 
fact, for $U_{,v}$ this divergence rate is never realized, because 
$U_{,w}$ always vanishes along the line $v=0$, but this is not 
important for our discussion). In addition, as long as $V_{,w}$ is 
nonzero at $v=0$ (which is indeed the case, as we shall immediately 
show), $V_{,vv}$ is dominated by the last term in the right-hand side of 
Eq. (7):
$$V_{,vv}\cong -\left[ {n^{-2}(n-1)V_{,w}} \right]_{}\kern 1pt |v|^{1/ n-
2}\;.\eqno(22)$$ 
At $u=u_0$, $V_{,w}$ is nothing but ${}^vV_{,w}$, which is given in Eq. 
(13).
\footnote {Despite the difference in the context, ${}^vV(v)$ is 
essentially the same as $V(v)$ of section \ref{sec 3}, and, more specifically, 
their $v$-derivatives (and hence also $w$-derivatives) are identical. 
To verify this, note that ${}^vV_{,v}$ is uniquely determined, through Eq. 
(19a), from the initial function ${}^vU(v)$ - in the same way that in the 
context of section \ref{sec 3} $V_{,v}$ is uniquely determined from $U(v)$ 
through Eq. (2). Since ${}^vU(v)$ is identical to the function $U(v)$ in Eq. 
(12), it follows that ${}^vV_{,v}$ is identical to $V_{,v}$ of Eq. (13).}
 Therefore, at $u=u_0$ we have 
$$V_{,w}(v=0)=2a(n-2)\ne 0\quad \quad (u=u_0)\;.$$	
From the smoothness of $V(u,w)$ it then follows that 
$$V_{,w}(v=0)\ne 0 \eqno(23)$$	
in a neighborhood of $u=u_0$. In this neighborhood, $V_{,vv}$ thus 
diverges like $\kern 1pt |v|^{1/ n-2}$. 
	
The Riemann component
$$R_{vxvx}={1 \over 2}e^{V-U}\left( {-V_{,vv}+U_{,v}V_{,v}-
M_{,v}V_{,v}} \right) \eqno(24)$$ 	
is dominated at $v=0$ by $V_{,vv}$:
$$R_{vxvx}\cong -{1 \over 2}e^{(V-U)}V_{,vv}\cong {1 \over 2}\left[ 
{n^{-2}(n-1)e^{(V-U)}V_{,w}} \right]\,\,|v|^{1/ n-2} \eqno(25)$$
(see footnotes 1,2 above). Along an outgoing null geodesic of constant 
$x,y,u,$ it is convenient to use the parallelly-propagated tetrad
$$e_x^\alpha =e^{(U-V)/ 2}\kern 1pt \delta _x^\alpha \;\;,\;\;e_y^\alpha 
=e^{(U+V)/ 2}\kern 1pt \delta _y^\alpha \;\;,\;\;e_u^\alpha =\sqrt 
2\,\delta _u^\alpha \;\;,\;\;e_v^\alpha =\sqrt 2\,e^M\kern 1pt \delta 
_v^\alpha $$
to evaluate the parallelly-propagated Riemann component
$$e_v^\alpha \,e_x^\beta \,e_v^\gamma \,e_x^\delta \,R_{\alpha \beta 
\gamma \delta }=2e^{(U-V+2M)}\kern 1pt R_{vxvx}\cong \left[ {n^{-
2}(n-1)e^{2M}V_{,w}} \right]\,\,|v|^{1/ n-2}\;.\eqno(26)$$
From Eq. (23) and the continuity of $M$ it is obvious that at $v=0$ the 
term in brackets is nonvanishing in a neighborhood of $u=u_0$. In this 
neighborhood, the parallelly-propagated Riemann component (26) 
diverges like $|v|^{1/ n-2}$.
	
We find that in the perturbed spacetime, too, the hypersurface $v=0$ is 
a null curvature singularity. This singularity is weak, because $U$, $V$, 
and $M$ are all finite at $v=0$. We shall now show that this singularity 
is scalar-curvature. To that end, we shall calculate the scalar $K\equiv 
R_{\alpha \beta \gamma \delta }R^{\alpha \beta \gamma \delta }$. 
Although the full expression for $K$ is fairly complicated, a 
straightforward calculation shows that near $v=0$ it is dominated by
$$K\cong 32\,\left[ {e^{2(M+U-
V)}R_{vxvx}R_{uxux}+e^{2(M+U+V)}R_{vyvy}R_{uyuy}} \right]\,\;.$$
This is the only part of $K$ that includes second-order $v$-derivatives, 
and it diverges like $v^{1/ n-2}$ (see below), whereas all other parts 
diverge like $v^{2/ n-2}$ or slower. One can easily verify that the two 
terms in the brackets are equal, and that $R_{uxux}$ is of the same 
form as $R_{vxvx}$ [Eq. (24)], except that $v$ is replaced by $u$. 
Expressing the dominant part of $R_{vxvx}$ in terms of $V_,vv$, using 
Eq. (25), we obtain
$$K\cong HV_{,vv}$$	
where
$$H\equiv 16\,e^{2M}\,\left( {V_{,uu}-U_{,u}V_{,u}+M_{,u}V_{,u}} 
\right)\;.$$ 	
In appendix A we show that for a generic choice of initial functions 
${}^uU(u)$ and $\Delta U(v)$, $H$ is nonvanishing at $v=0$ in a 
neighborhood of $u=u_0$. In view of Eqs. (22,23), the scalar $K$ 
diverges there like $v^{1/ n-2}$. The hypersurface $v=0$ is thus a 
scalar-curvature singularity.
	
Let us summarize the main results of this section. The perturbed 
spacetime (described by a CPW solution) has the following features:
\newline
i)   No new singularity (spacelike or whatsoever) forms in $S^-$ (i.e. 
before $v=0$);
\newline
ii)   The hypersurface $v=0$ remains a null, weak, curvature 
singularity;
\newline
iii)   The divergence rate of the most divergent PP Riemann components 
is unchanged: $v^{1/ n-2}$.
\newline
iv)   For a generic outgoing perturbation, however, the singularity 
becomes scalar-curvature.

\section{Discussion}   \label{sec5}

We have shown that within the framework of (linearly-polarized) 
plane-symmetric solutions the null weak singularity (4,5) is stable to 
both ingoing and outgoing perturbations: Both types of perturbations 
preserve the null weak character of the singularity. However, the 
outgoing perturbations generically convert the original non-scalar PP 
curvature singularity into a scalar-curvature singularity. This behavior 
is compatible with what we know about the CH singularity in black 
holes: In the mass-inflation model \cite{12}, outgoing radiation 
converts the non-scalar PP singularity of the charged Vaidya solution 
\cite{16,27} into a scalar-curvature singularity. Note also that the CH 
singularity in a generic spinning black hole is scalar-curvature. 
\cite{10,23}
	
The local stability of null weak singularities, demonstrated here (and 
also in Ref. \cite{25}), provides a strong support to the new picture of 
the CH singularity, which was obtained primarily from the perturbative 
approach \cite{10,11}. The present model of null weak singularities in 
CPW spacetimes may also serve as a useful toy-model for analysing 
various features of null weak singularities.
	
Our analysis rules out the possibility that the introduction of outgoing 
perturbations will transform the entire null singularity into a 
spacelike one. It is still possible that, as the result of the outgoing 
perturbations, the null singularity will terminate at some $u=u_s>u_0$, 
where it intersects a spacelike singularity. This would be consistent 
with our construction, because, in the analysis in section \ref{sec 4}, the 
demonstration of the regularity of $U(u,w)$ is restricted to the region 
$S$, i.e. $u\le u_2$. It thus may be possible that a spacelike singularity 
will be present at $u>u_2$. In fact, a spacelike singularity will 
positively form if $W$ vanishes at some $u_0<u<u_1$, as $U=-\ln (W)$ 
will diverge there. (Our analysis only guarantees that this cannot 
happen in the neighborhood of the initial hypersurface $u=u_0$. Note 
also that if the outgoing perturbation is bounded in a suitable way, then 
such a divergence will be excluded in the entire range $u_0<u<u_1$.) 
The line $W(u,v)=0$ (when exists) is known to be the locus of a 
spacelike, asymptotically Kasner-like, curvature singularity \cite{22}. 
Note that this situation of a null singularity becoming spacelike at a 
certain point also occurs in the model of a spherically-symmetric 
charged black hole perturbed by a self-gravitating scalar field 
\cite{15}. (At present, however, it is unclear whether such a situation 
also occurs in vacuum spinning black holes.)
	
The solutions constructed here are all of the asymptotic form 
$V\propto v^{1/ n}$ for some constant $n>2$. This is also the situation 
in the analysis by OF \cite{25} - except that in the latter, unlike here, 
$n$ was an integer. We shall refer to this type of asymptotic behavior 
as the {\it power-law} asymptotic behavior. The asymptotic behavior at the 
CH singularity of realistic black holes (as emerges from the 
perturbation analyses) is somewhat different, as we now explain:
\newline
*  For {\it axially-symmetric} perturbations of a Kerr BH \cite{10} (and also 
for a perturbed RN BH \cite{13}), the asymptotic behavior is $V\propto 
(\ln v)^{-n}$, where $n$ is the integer characterizing the power-law 
tails. We shall refer to it as the {\it logarithmic} asymptotic behavior.
\newline
*  For {\it nonaxially-symmetric} perturbation modes of a Kerr BH, the 
asymptotic behavior is $V\propto (\ln |v|)^{-n}\,\cos (m\omega \;\ln 
|v|)$, where $\omega $ is a constant and m is the magnetic number of 
the mode in question. \cite{10} We shall refer to it as the 
{\it oscillatory-logarithmic} asymptotic behavior.
\newline
[Here $v$ is the affine parameter along an outgoing null geodesic, with 
$v=0$ at the CH singularity, and $V$ stands for a typical metric 
function (in a suitable gauge).] One would certainly like to extend the 
present analysis to these more realistic types of asymptotic behavior. 
It seems that the generalization to the logarithmic asymptotic 
behavior will be almost straightforward if one replaces $w\equiv 
|v|^{1/ n}$ of the present analysis by $w\equiv (\ln |v|)^{-n}$. The 
generalization to the oscillatory-logarithmic case is less 
straightforward, because of its non-monotonic nature. [In particular, 
when analysing Eq. (16b), it will not be possible to use $w\equiv (\ln 
|v|)^{-n}\,\cos (\omega \;\ln |v|)$ as a coordinate instead of $v$.] Still, 
due to the simplicity of the equations describing CPW spacetimes, 
hopefully the generalization to this type of asymptotic behavior will 
not be too difficult. Note that the approach used by OF \cite{25} seems 
to be inapplicable even for the logarithmic asymptotic behavior, 
because it is based on analyticity: Whereas for $w\equiv (-v)^{1/ n}$ 
(with integer $n$) $v$ is an analytic function of $w$, for $w\equiv (\ln 
|v|)^{-n}$ $v(w)$ fails to be analytic at $v=0$. For the same reason, in 
Ref. \cite{25} the power index $n$ had to be an integer, whereas here 
$n$ can be any real number $>2$. In this respect, the present approach 
has an advantage over that of OF (however, as we mentioned in the 
introduction, in other respects the analysis by OF yields much more 
powerful results).
	
The present construction was restricted to $n>2$. It is 
straightforward, however, to extend it to any positive $n\ne 1,2$. The 
features of the resultant singularity at $v=0$ will significantly depend 
on the value of $n$. This issue deserves further investigation. It will 
also be interesting to generalize the present analysis to the 
logarithmic and oscillatory-logarithmic cases described above, and to 
the arbitrarily-polarized case.

\section*{Appendix  A}

Let us define
$$H_0(v)\equiv \left( {V_{,uu}-U_{,u}V_{,u}+M_{,u}V_{,u}} 
\right)_{u=u_0}\;.$$	
It is sufficient to show that 
$$H_0(v=0)\ne 0\;.\eqno(A1)$$
Then, $H(u=u_0\,,\,v=0)=16\,e^{2M}H_0(v=0)$ is nonvanishing too; and 
from continuity $H(v=0)$ is nonzero in some neighborhood $u>u_0$.
	
In order to calculate $H_0(v)$ we must evolve the $u$-derivatives of 
$U$,$V$, and $M$ along the line $u=u_0$, from $v=v_0$ (where these 
derivatives are obtained directly from the $u$-dependent initial 
functions), and up to $v=0$. To that end we shall use the evolution 
equations (16a-c), which may be viewed as ordinary differential 
equations for the $u$-derivatives of the metric functions. Let us define
$$U_u(v)\equiv U_{,u}(u=u_0,v)\;,\;V_u(v)\equiv 
V_{,u}(u=u_0,v)\;,\;M_u(v)\equiv M_{,u}(u=u_0,v)$$	
and
$$V_{uu}(v)\equiv V_{,uu}(u=u_0,v)\;,$$
so
$$H_0=V_{uu}-U_uV_u+M_uV_u\;.\eqno(A2)$$
From Eqs. (20,21) it follows that
$$U_u=\exp \left[ {{}^vU(v)-^0U} \right]\,^uU_{,u}(u_0)\;,$$ 
where ${}^0U\equiv {}^vU(v=v_0)={}^uU(u=u_0)$. For our purpose, it is 
sufficient to recall that the functional dependence of $U_u$ on the 
initial functions is
$$U_u\equiv U_u\left[ {{}^vU\,;\,^uU_{,u}(u_0)} \right]\;.\eqno(A3)$$ 
Here and below, the semicolon distinguishes between the functions of 
$v$ (i.e. the initial functions at $u=u_0$) at the left, and the 
parameters (obtained by evaluating the $u$-dependent initial functions, 
and/or their derivatives, at $u=u_0$) at the right. (For brevity, we omit 
the obvious dependence on $v$ from this list of dependencies.) Next, 
applying Eq. (16b) to $u=u_0$, we obtain
$$V_{u,v}={\textstyle{1 \over 2}}(U_u{}^vV_{,v}+{}^vU_{,v}V_u)\;.$$
$V_u(v)$ is to be determined from this ordinary differential equation, 
together with the initial condition $V_u(v=v_0)={}^uV_{,u}(u_0)$. It is 
straightforward to solve this linear equation explicitly (recall that 
$U_u$, ${}^vU$, and ${}^vV_{,v}$ are known), but again, for our purpose it 
is sufficient to recall that the functional dependence of $V_u$ on the 
initial functions is
$$V_u\equiv V_u\left[ {U_u,\,{}^vV_{,v},\,{}^vU_{,v}\,;\,^uV_{,u}(u_0)} 
\right]\equiv V_u\left[ {{}^vU\,;\,^uV_{,u}(u_0),\,^uU_{,u}(u_0)} 
\right]\;.\eqno(A4)$$ 
[Here and below, we take into consideration Eq. (19a), which allows us 
to express ${}^vV_{,v}$ in terms of the function ${}^vU(v)$]. In a similar 
manner, applying Eq. (16c) to $u=u_0$, we obtain
$$M_{u,v}={\textstyle{1 \over 2}}({}^vV_{,v}V_u-{}^vU_{,v}U_u)\;.$$ 
$M_u$ can thus be obtained by a direct integration, meaning that its 
functional dependence is
$$M_u\equiv M_u\left[ {V_u,U_u,\,{}^vV_{,v},\,{}^vU_{,v}\,} \right]\equiv 
M_u\left[ {\,{}^vU\,;\,^uV_{,u}(u_0),\,^uU_{,u}(u_0)} \right] \eqno(A5)$$
(recall that $M_u$ vanishes at $v=v_0$). In summary, the three 
functions $U_u$, $V_u$ and $M_u$ are fully determined from the 
function ${}^vU$ and the two parameters $^uV_{,u}(u_0)$, 
$^uU_{,u}(u_0)$.
	
Let us now analyse $V_{uu}$. Differentiating Eq. (16b) we find
$$V_{uu,v}\equiv \left[ {(V_{,uv})_{,u}} \right]_{u=u_0}={1 \over 
2}\left( {U_{,uu}V_{,v}+U_{,v}V_{,uu}+U_{,u}V_{,uv}+U_{,uv}V_{,u}} 
\right)_{u=u_0}\;.$$
 Using Eq. (17b), we obtain a differential equation of the form
$$V_{uu,v}={1 \over 2}{}^vU_{,v}V_{uu}+{1 \over 2}G(v) \eqno(A6)$$
where
$$G(v)\equiv \left[ \left( {U_u}^2+{V_u}^2 \right)/ 2-M_uU_u 
\right]\kern 1pt {}^vV_{,v}+U_uV_{u,v}+U_{u,v}V_u\;.$$ 
Note the functional dependence of $G(v)$,
$$G(v)=G\left[ {V_u,U_u,M_u,\,{}^vV_{,v}\,} \right]=G\left[ 
{{}^vU\,;\,^uV_{,u}(u_0),\,^uU_{,u}(u_0)} \right]\;.$$	
Equation (A6) is a linear ordinary differential equation for $V_{uu}(v)$, 
whose general solution is
$$V_{uu}(v)=\exp \left( {{}^vU/ 2} \right)\;\left( {C+{1 \over 
2}\int\limits_{v_0}^v {G(v')\exp \left[ {-{}^vU(v')/ 2} \right]dv'}} 
\right)\;,$$
where $C$ is an integration constant. Recalling the initial condition,
$$V_{uu}(v=v_0)={}^uV_{,uu}(u=u_0)\;,$$
we find
$$V_{uu}(v)=Z(v)+\exp \left[ {({}^vU-{}^0U)/ 2} 
\right]\,^uV_{,uu}(u_0)\;,\eqno(A7)$$ 
where 
$$Z(v)\equiv {1 \over 2}\exp ({}^vU/ 2)\int\limits_{v_0}^v {G(v')\exp 
\left[ {-{}^vU(v')/ 2} \right]dv'}\;,$$
namely
$$Z(v)=Z\left[ {{}^vU\,,G} \right]=Z\left[ 
{{}^vU\,;\,^uV_{,u}(u_0),\,^uU_{,u}(u_0)} \right]\;.$$
	
Collecting Eqs. (A3,A4,A5,A7) and substituting in Eq. (A2), we can 
reexpress $H_0$ as
$$H_0(v)=Q(v)+\exp \left[ {({}^vU-{}^0U)/ 2} \right]\,^uV_{,uu}(u_0)\;,$$ 
where 
$$Q(v)\equiv Z(v)-U_uV_u+M_uV_u$$
has the functional dependence
$$Q=Q\left[ {{}^vU\,;\,^uV_{,u}(u_0),\,^uU_{,u}(u_0)} \right]\;.$$ 
At $v=0$, we find 
$$H_0(v=0)=q_0+\exp \left( {\left[ {{}^vU(v=0)-{}^0U} \right]/ 2} 
\right)\,^uV_{,uu}(u_0) \eqno(A8)$$	
where
$$q_0\equiv Q(v=0)=q_0\left[ {\kern 1pt 
{}^vU\,;\,^uV_{,u}(u_0),\,^uU_{,u}(u_0)} \right]\;.$$
	
We still need to express the parameters $^uV_{,u}(u_0)$ and 
$^uV_{,uu}(u_0)$ in terms of the initial function $^uU(u)$, through Eq. 
(19b). The latter implies
$$^uV_{,u}=\pm \left( {2\,^uU_{,uu}-{^u{U_{,u}}^2}} \right)^{1/ 
2}\;\;,\;\;^uV_{,uu}=\pm \left( {\,^uU_{,uuu}-{^uU_{,u}}\;^uU_{,uu}} 
\right)\left( {2\,^uU_{,uu}-{^u{U_{,u}}^2}} \right)^{-1/ 2}\;,$$
so we can reexpress the functional dependence of $q_0$ as 
$$q_0=q_0\left[ {{}^vU\,;\,^uU_{,u}(u_0),\,^uU_{,uu}(u_0)} \right]\;.$$ 
We can now write Eq. (A8) in the form
$$H_0(v=0)=h_0+h_1\,^uU_{,uuu}(u_0)\;,\eqno(A9)$$	
where
$$h_0=q_0\pm \exp \left[ {[{}^vU(v=0)-{}^0U]/ 2} \right]\;\left( {-
{^uU_{,u}}(u_0)^uU_{,uu}(u_0)} \right)\left( {2\,^uU_{,uu}(u_0)-
{^uU_{,u}(u_0)^2}} \right)^{-1/ 2}$$
and
$$h_1=\pm \exp \left[ {[{}^vU(v=0)-{}^0U]/ 2} \right]\,\,\left[ 
{2\,^uU_{,uu}(u_0)-\,^uU_{,u}(u_0)^2} \right]^{\kern 1pt -1/ 2}$$ 
are two numbers with the same type of functional dependence:
$$h_{0,1}=h_{0,1}\left[ {{}^vU\,;\,^uU_{,u}(u_0),\,^uU_{,uu}(u_0)} 
\right]\;.$$ 	
Obviously, $h_1\ne 0$. Therefore, in view of Eq. (A9), for any choice of 
initial function ${}^vU(v)$ and parameters 
$^uU_{,u}(u_0)\,,\,^uU_{,uu}(u_0)$ there only exists a single value of 
$^uU_{,uuu}(u_0)$ for which $H_0(v=0)=0$. Thus, for a generic choice of 
the initial function $^uU(u)$, the inequality (A1) is satisfied.

This research was supported in part by the United States--Israel 
Binational Science Foundation, and by the Fund for the Promotion of 
Research at the Technion.



\begin{thebibliography}{99}
\bibitem{24}  This is probably the reason why the type of CPW 
solutions constructed here, with a locally-stable null weak singularity, 
was left unnoticed so far. In previous analyses of singularities in CPW 
spacetimes, attention was restricted to global solutions arising from 
regular initial data.
\end{thebibliography}
\end{document}